\documentclass[preprint,12pt]{elsarticle}

\usepackage{amssymb}

\begin{document}

\begin{frontmatter}

\title{On The Problem of Vacuum energy in Brane Theories}

\author{Ilya Gurwich}
\ead{gurwichphys@gmail.com}
\author{Shimon Rubin}
\ead{rubinsh@bgu.ac.il}
\author{Aharon Davidson}
\ead{davidson@bgu.ac.il}
\address{Physics Department, Ben-Gurion University,
Beer-Sheva 84105, Israel}

\begin{abstract}
We point out that modern brane theories suffer from a severe vacuum
energy problem.
To be specific, the Casimir energy associated with the matter fields
confined to the brane, is stemming from the one and the same
localization mechanism which forms the brane itself, and is thus
generically unavoidable.
Possible practical solutions are discussed, including in particular
spontaneously broken supersymmetry, and quantum mechanically
induced brane tension.
\end{abstract}


\end{frontmatter}

The idea that our universe is a brane embedded in a higher
dimensional space-time has received a great deal of attention
for several reasons.
First and foremost, quantum gravity seems to demand it, and
to that argument joins superstring/M theory which predicts
ten/eleven dimensions of space-time.
Brane gravity has made some remarkable progress over the last
few years, dynamical localization mechanisms have been found,
and many 4-dim general relativistic results have been
reproduced\cite{RS,DGP,CH,RSgr}.

The most fundamental and important aspect of brane theory is
that although we live in a high dimensional space (the bulk),
all Standard Model fields are localized on a 4-dim hypersurface
(the brane) with some finite thickness $\delta$.
This brane thickness is often taken to be zero for simplifying
calculations, but in all realistic models, especially those which
include quantum corrections, this thickness must be finite.
The limits on 4-dim gravity at low scales are fairly loose.
We know that gravity is 4-dim to about 10 microns, and
different brane models make use of that loose limit.
Standard Model fields, however, are much more confined.
And since no accelerator ever detected signatures that can
be interpreted as higher dimensional propagation, one deduces
that $\delta \leq (1TeV)^{-1}$.
This upper bound is completely independent of the brane model
at hand, and will be the main source of the issue discussed
in this Letter.
To be precise, it will be translated into a lower bound on the
"residual" vacuum energy on the brane, to be regarded as an
unavoidable outcome of brane gravity.

The vacuum energy problem is still an open question.
According to quantum field theory (QFT), summing the
zero-point energies of all normal modes of matter fields up to
the Planck scale (or even the QCD cutoff) gives  rise to enormous
energy density of the vacuum around us.
Despite this, no such energy density seems to exist 
(dark energy resembles such an energy density, but its observed
value is inexplicably smaller than QFT predicts).
However, we also know that vacuum energy does exist in some
form because effects originated from vacuum fluctuations have
been predicted and measured.
One such effect, and perhaps the most direct observation
of vacuum energy, is the Casimir effect \cite{CAStheo,CASexp}.
In general terms, the Casimir effect is the variation in vacuum
energy caused by the addition of boundary conditions to the
system.
A quantum field subject to boundary conditions (caused by other
matter fields or strongly curved space-time) would have a different
vacuum energy than a free field.
The difference between the vacuum energies of the constrained
and free field is the Casimir energy (note that the Casimir
energy is independent of the QFT cutoff).
In fact, it is the quantum backreaction of the field to the
boundary conditions.
A concrete example of such effect is the attractive force between
two parallel conductive plates in a vacuum.
The plates create boundary conditions for the electro-magnetic
field, and thus a force is generated \cite{CAStheo}.

\begin{figure}[ht]
  \centering
	\includegraphics[scale=0.75]{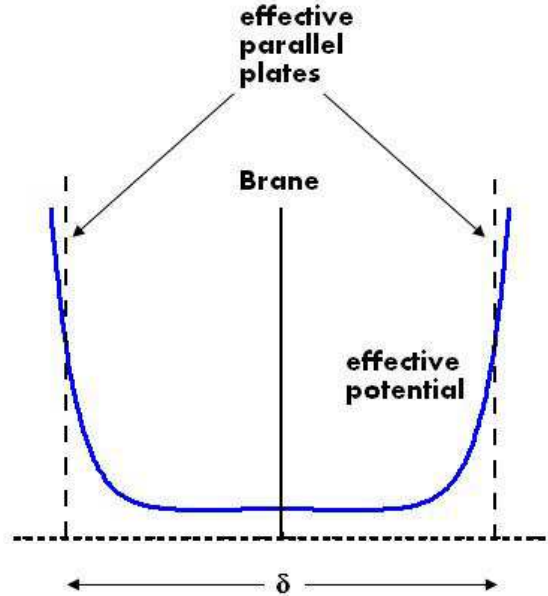}
	\caption{\label{fig1}Effective potential: In brane scenario,
	the Standard Model fields are localized around a single brane.
	This can be described via an effective localization potential
	with a very sharp minima. The latter can be approximated
	by a potential well with a certain width $\delta$.}
\end{figure}

In brane theory this means that the localization of standard
model fields on the brane, regardless of the underlying
mechanism\cite{brloc}, results in essentially the same effect
as the one caused by a pair of conducting plates (Fig.\ref{fig1}).
Several papers dealing with the Casimir effect in brane gravity
have already been presented\cite{BulkCAS,BraneCAS,BraneCASstabil,BraneSUSYCAS},
but most of them focus on the Casimir effect between two branes.
Unlike the latter, the present work is model independent, and
deals with the Casimir effect generated by the familiar matter
fields embedded on a single brane.
Once these fields get localized to the brane, their confinement
is analogous to the confinement of an electro-magnetic field
between conducting plates.
The form of the exact localization mechanism is unimportant,
as it will always result in the same energy up to a constant of
${\cal O}(1)$.

The Casimir energy of two plates in 4+n dimensions with a
separation $\delta$ \footnote[1]{We assume all Standard Model
fields are localized in the same way.
We also assume for simplicity, that all masses are much lower
than the localization scale $\delta^{-1}$, so we may treat
them as massless. both assumptions may not be exact but they
greatly simplify the equations and deviating from these
assumptions will not modify our conclusions at all.},
is given by \cite{CAStheo}
\begin{equation}
	E=\frac{\eta\hbar A}{\delta^{3+n}} ~,
	\label{casimir}
\end{equation}
where $A$ is the area of the plates (hyper area, in the general
case) and $\eta$ is a constant.
From this point on, we will use the notation $\hbar=1$.
The value of $\eta$ depends on $n$, on the exact form
of the localization potential, and on the number of standard
model fields (degrees of freedom).
The energy density between the plates is therefore
\begin{equation}
	\rho_{bulk}=\frac{\eta}{\delta^{4+n}} ~.
	\label{casimir_dens}
\end{equation}

Since we are talking about a D3 brane, and assume more than
one extra-dimension, we do not face a parallel plate system,
but rather, cylindrical or spherical boundaries.
While leaving the form of eq.(\ref{casimir_dens}) intact,
this will accordingly modify the value of $\eta$.
Eq.(\ref{casimir_dens}) describes the bulk energy density.
The energy density on the brane is obtained by integrating
out all $n$ extra-dimensions, so that
\begin{equation}
	\rho=\frac{\tilde{\eta}}{\delta^{4}} ~,
	\label{brane_dens}
\end{equation}
with $\tilde{\eta}\neq \eta$.
This is a \textbf{constant} energy density on the brane, thus
it is a direct contribution to the cosmological constant
\footnote[2]{We note that brane theories can bring forward
other contributions to the cosmological constant. However,
the Casimir energy is the only model independent contribution.}.
The exact value of $\tilde{\eta}$ is of course model dependent,
but it cannot deviate too much from ${\cal O}(1)$.
With this in mind, taking into account the experimental bound
on $\delta$, we evaluate the energy density from
eq.(\ref{brane_dens}) and find
\begin{equation}
	\rho\geq\left(1TeV\right)^4 ~.
	\label{brane_dens_num}
\end{equation}
This is to be contrasted with the much smaller value of the
cosmological constant (dark energy)
$\rho_{\Lambda}\sim\left(10^{-3}eV\right)^4$, leading to
a 60 orders of magnitude discrepancy.
Unlike the ordinary vacuum energy, this energy cannot be
'swept under the carpet' because it does not stem directly
from the action, but rather, caused by (quantum corrections
due to) the abnormal structure of matter and space-time.
This constitutes a serious problem.
In order for brane theories to be realistic, one must find a
way to cancel or suppress this energy.

It is natural to turn first to supersymmetry, the natural cure
for vacuum energy \cite{SUSY}.
If unbroken, SUSY assures an absolute cancellation of the
vacuum energy.
However, we know that SUSY must be broken at an energy
scale $M_{SUSY}\geq1TeV$.
If SUSY is broken at a much lower energy scale than the
localization energy, then we might expect a strong
suppression of the Casimir energy \cite{SUSYCAS}.
As a simple example, let us consider the case of a scalar
field and its superpartner "calar".
In order to perform exact calculations, we assume one
extra-dimension ($n=1$) and simplify the localization 
potential to be an infinite well of width $\delta$.
The Casimir energy density generated by the scalar field
on the brane is given by\cite{CAStheo}
\begin{equation}
	\rho_{scalar}=-2\left(\frac{m}{4\pi}\right)^{5/2}
	\frac{1}{\delta^{3/2}}\sum^{\infty}_{j=1}
	\frac{K_{5/2}\left(2m\delta j\right)}{j^{5/2}} ~,
	\label{casimir_scalar}
\end{equation}
where $K_{\nu}(x)$ is the modified Bessel function of the second
type, and $m$ is the mass of the scalar field.
If SUSY is unbroken, the "calar" field will give the exact same
result but with an opposite sign.
However, if SUSY is broken, even at an energy scale lower
than $\delta^{-1}$, then the masses will be slightly corrected,
such that $m^{2}_{scalar}-m^{2}_{calar}=\Delta m^2$.
In that case, and assuming $M_{SUSY}>>m>>\Delta m$,
the residual Casimir energy density on the brane becomes
\begin{equation}
	\rho\cong lim_{m\rightarrow0}\frac{\Delta m^2}{2m}
	\frac{\partial\rho_{scalar}}{\partial m} ~,
	\label{casimir_susy1}
\end{equation}
where $\rho=\rho_{scalar}+\rho_{calar}$.
Evaluating eq.(\ref{casimir_susy1}) using eq.(\ref{casimir_scalar}),
we obtain
\begin{equation}
	\rho=\frac{\zeta(3)}{64\pi^2}
	\left(\frac{\Delta m}{\delta}\right)^2~.
	\label{casimir_susy2}
\end{equation}
Thus, we obtain a suppression factor of
$\left(\Delta m\delta\right)^2$ to the original (non-SUSY)
Casimir energy.
The form of $\Delta m$ dependents on the mechanism by
which SUSY is broken.
If it is spontaneously broken \cite{SUSYsb}, it will usually
take the form
\begin{equation}
	\Delta m\sim m e^{-M_{LSM}/M_{SUSY}} ~,
	\label{delta_m}
\end{equation}
where $M_{LSM}=\delta^{-1}$ is the localization energy
scale of the Standard Model fields.
In turn, eq.(\ref{casimir_susy2}) takes the form
\begin{equation}
	\rho=\frac{\zeta(3)}{64\pi^2}
	\left(M_{LSM}m e^{-M_{LSM}/M_{SUSY}}\right)^2~.
	\label{casimir_susy3}
\end{equation}
Thus, we need at least $M_{LSM}/M_{SUSY}\sim100$ to
account for the discrepancy between the Casimir energy
and the observed cosmological constant.
In other words, we must have a clear hierarchy
between the SUSY scale and the extra-dimensional scale.

There is another way to deal with the excess of Casimir energy.
In the Randall-Sundrum scenario, as well as the subsequent
Collins-Holdom and unified brane gravity, the Casimir energy
can in fact play the role of the brane surface tension\cite{RS}
\begin{equation}
	\sigma=\frac{1}{4\pi G_5}\sqrt{\frac{-\Lambda_5}{6}} ~,
	\label{RS_ST}
\end{equation}
where $G_5$ is the bulk gravitational constant and $\Lambda_5$
is the bulk cosmological constant.
As far as energy scales are concerned, $1/G_5=4\pi M^{3}_{5}$,
where $M_5$ is the bulk Planck mass, and
$\sqrt{-\Lambda_5/6}=M_{LG}$ is the localization energy of gravity.
If we want the Casimir energy to be the source of the surface tension,
we need
\begin{equation}
	\tilde{\eta}M^{4}_{LSM}=M^{3}_{5}M_{LG} ~.
	\label{Casimir_RS_ST}
\end{equation}
This is obtained by comparing eq.(\ref{RS_ST}) and
eq.(\ref{brane_dens}).
Note that this solution is valid only for $\tilde{\eta}>0$.
All the three scales which govern eq.(\ref{Casimir_RS_ST})
can in principle be measured experimentally.
$M_{LSM}$, as we already noted, is the energy at which
colliders should detect propagation of particles in
extra-dimensions, since particles above that energy are no longer
localized to the brane.
$M_5$ is the energy scale at which black holes should start
forming in colliders, since this is the scale of quantum gravity
\cite{LHCbh}.
And $M_{LG}$ is measurable via small-scale gravity experiments
that are also conducted in an advancing rate \cite{SLSgrav}.
For scales below $M^{-1}_{LG}$, gravity no longer behaves
4-dimensionally and the gravitational force no longer behaves
as $1/r^2$ but rather as $1/r^{2+n}$.
Therefore, eq.(\ref{Casimir_RS_ST}) gives a unique prediction
for the relation between the various scales involved.
This prediction can be translated into a fine-tuning problem.
To solve this fine-tuning, one would presumably need a
dynamical mechanism to determine one of the bulk or brane
parameters.
Several works have been made to demonstrate how it is possible
to determine the bulk AdS scale dynamically, so that the brane
would be almost flat \cite{BraneCASstabil}.
Eq.(\ref{Casimir_RS_ST}) holds for a wide range of models with
an AdS bulk.
For the original Randall-Sundrum model, that has no induced
gravity term, we have in addition, the relation connecting
the bulk constants to the 4-dim Planck mass $M_4$\cite{RS}
\begin{equation}
	M^{3}_{5}=M^{2}_{4}M_{LG}~.
	\label{RS_PM}
\end{equation}
In this case, we can make an even stronger prediction, since the
value of $M_4$ is known, namely
\begin{equation}
	\sqrt{\tilde{\eta}}M^{2}_{LSM}=M_{4}M_{LG}~.
	\label{Casimir_RS}
\end{equation}
Since we assume that all scales are lower than $M_4$, we see
from eq.(\ref{Casimir_RS}) that $M_{LG}<M_{LSM}$, that gravity
is less localized than matter fields.
Substituting this into eq.(\ref{Casimir_RS_ST}), we are led to
$M_5>M_{LSM}$.
This means, in contrast with current expectations from the
LHC, that black holes cannot be formed in colliders before
propagation of Standard Model particles into extra-dimensions
will be observed.

To summarize, we have shown that brane theories generically
suffer from an excess of vacuum energy of over 60 orders of
magnitude.
This vacuum energy originates from the Casimir effect, which
unlike the total vacuum energy, has been shown to be realized
in nature.
This problem is unavoidable because the Casimir energy stems
from the localization mechanism of the Standard Model fields,
the same mechanism that defines the brane in the first place.
We have suggested two possible ways to overcome this problem.
The first is the introduction of spontaneously broken SUSY.
If future measurements show that the hierarchy between the SUSY
breaking scale and the localization of
Standard Model fields scale is about two orders of magnitude,
then the SUSY suppressed Casimir energy can serve as the source
of dark energy.
The second idea was to utilize this energy to account for the brane
tension required in the Randall-Sundrum scenario
(and the related Collins-Holdom and unified brane gravity scenarios,
that also incorporate an AdS bulk). 
Each of these solutions requires a specific mass hierarchy relation.
The first solution dictates that the SUSY breaking scale is lower
than the Standard Model localization scale, and therefore 
SUSY effects will be observed prior to the propagation of particles
in extra-dimensions.
The second solution dictates a bulk Planck scale higher
than the Standard Model localization scale, and therefore black holes
will only be seen in colliders after the propagation in extra-dimensions.

\medskip
The authors would like to thank Ramy Brustein for useful comments
on SUSY topics and enlightening discussions.

\end{document}